\documentclass[12pt]{article}
\usepackage{epsfig}

\begin{document}

\begin{flushright}
Preprint PSI--PR--00--12 
\\[10mm]
\end{flushright}

\begin{center}
{\Large\sf The structure of the light scalar mesons and\\ 
  QCD sum rules}
\\[5mm]
{\sc V.E.~Markushin}
\\[5mm] 
{\it Paul Scherrer Institute, CH-5232 Villigen PSI, Switzerland}
\\[5mm]
{\small Presented at Meson 2000, Cracow, Poland, May 19--23, 2000\\
to be published in Acta~Phys.~Pol.~B }
\\[5mm]
\end{center}

\begin{abstract}

The structure of the light scalar mesons is elucidated by the investigation  
of the $S$-matrix poles and the $q\bar{q}$ spectral density in a coupled 
channel model that includes $\pi\pi$, $K\bar{K}$ and $q\bar{q}$ channels. 
It is shown that the dynamical origin of the $\sigma$ and $f_0(980)$ mesons 
is consistent with the observed spectrum of the scalar states.  
The $K\bar{K}$ molecular picture of the $f_0(980)$ is  
in good agreement with recent experimental data on the decay 
$\phi\to\gamma\pi\pi$ from Novosibirsk.  

\end{abstract}
  
\section{Introduction}

   The structure of the light scalar mesons addresses different problems  
in hadron spectroscopy: strong channel coupling, the OZI rule violation, 
exotic states (glueballs, $q^2\bar{q}^2$), and the properties of the QCD vacuum.    
   Recent theoretical studies (see \cite{To96,LMZ98,Ma99,KLL99} 
and references therein) have demonstrated that the appearance of dynamical 
states can naturally explain the striking difference of the observed spectrum 
of the $J^{PC}=0^{++}$ states \cite{RPP98} from the other multiplets.    
   The goal of this paper is to investigate the properties of the 
light scalar mesons in a coupled channel model where both the $\sigma$ meson 
and the $f_0(980)$ arise as dynamical states due to strongly attractive 
interactions in the $\pi\pi$ and $K\bar{K}$ channels.   

\begin{figure} 
\mbox{\epsfysize=120mm \epsffile{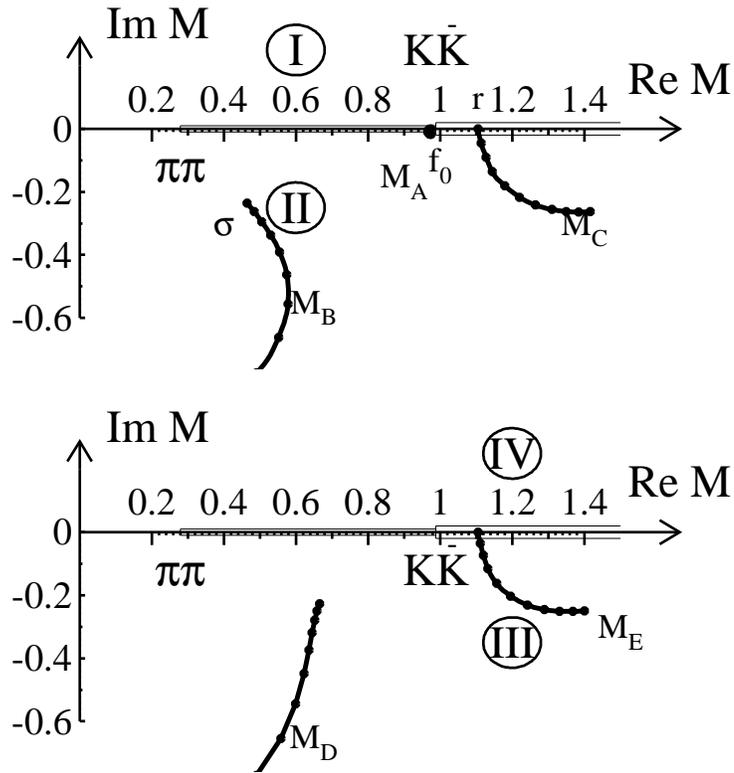}}
\caption{\label{Fig1}
The trajectories of the $S$-matrix poles in the complex mass plane (GeV)
on the sheets II and III  for the $K\bar{K}-\pi\pi$ and $q\bar{q}-\pi\pi$ 
couplings increasing from zero to the physical values.  
} 
\end{figure}

\begin{figure} 
\mbox{\hspace*{30mm}(a) \hspace*{60mm} (b)}\\[-5mm]
\mbox{
\mbox{\epsfxsize=60mm \epsffile{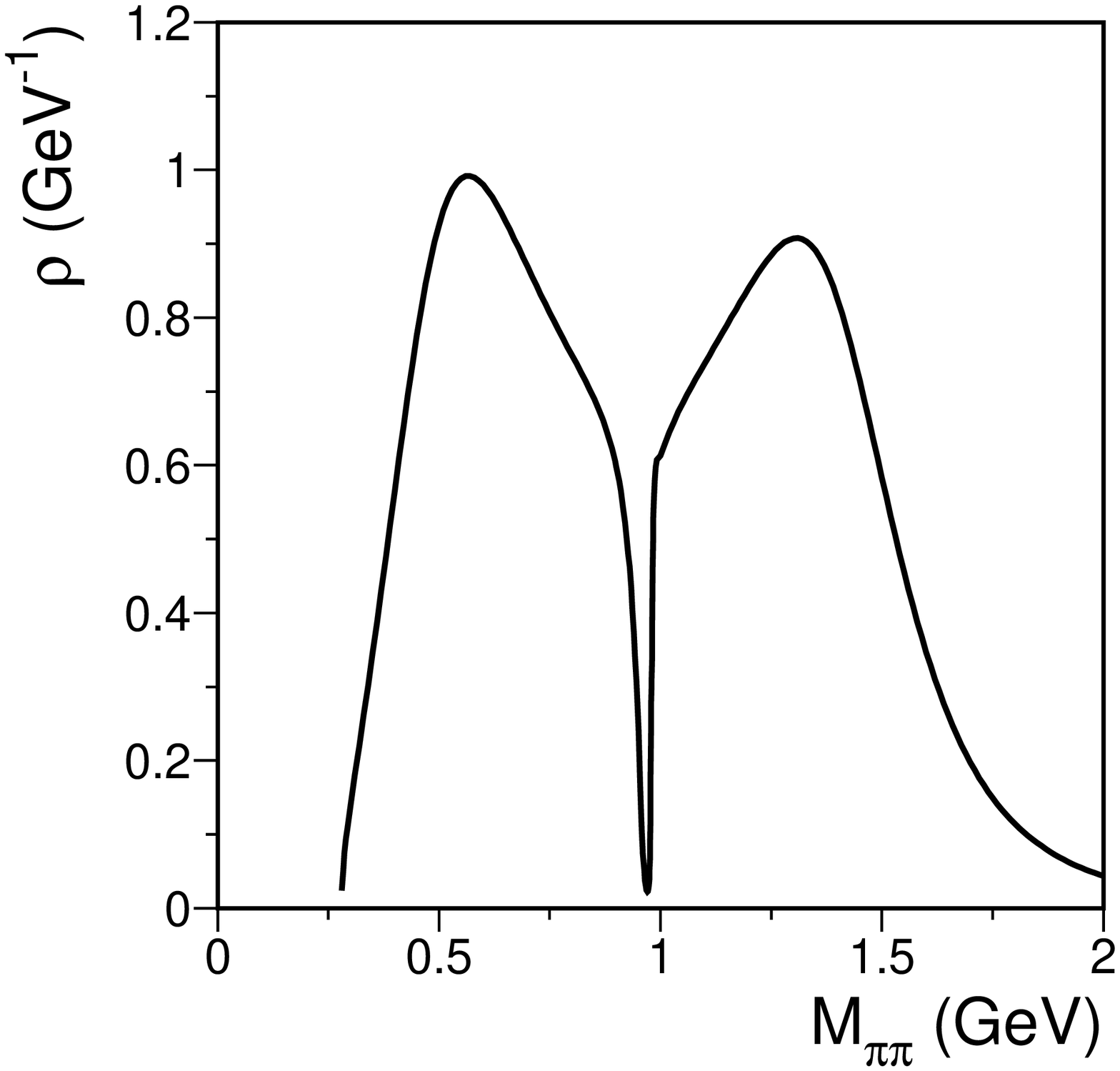}}
\mbox{\epsfxsize=60mm \epsffile{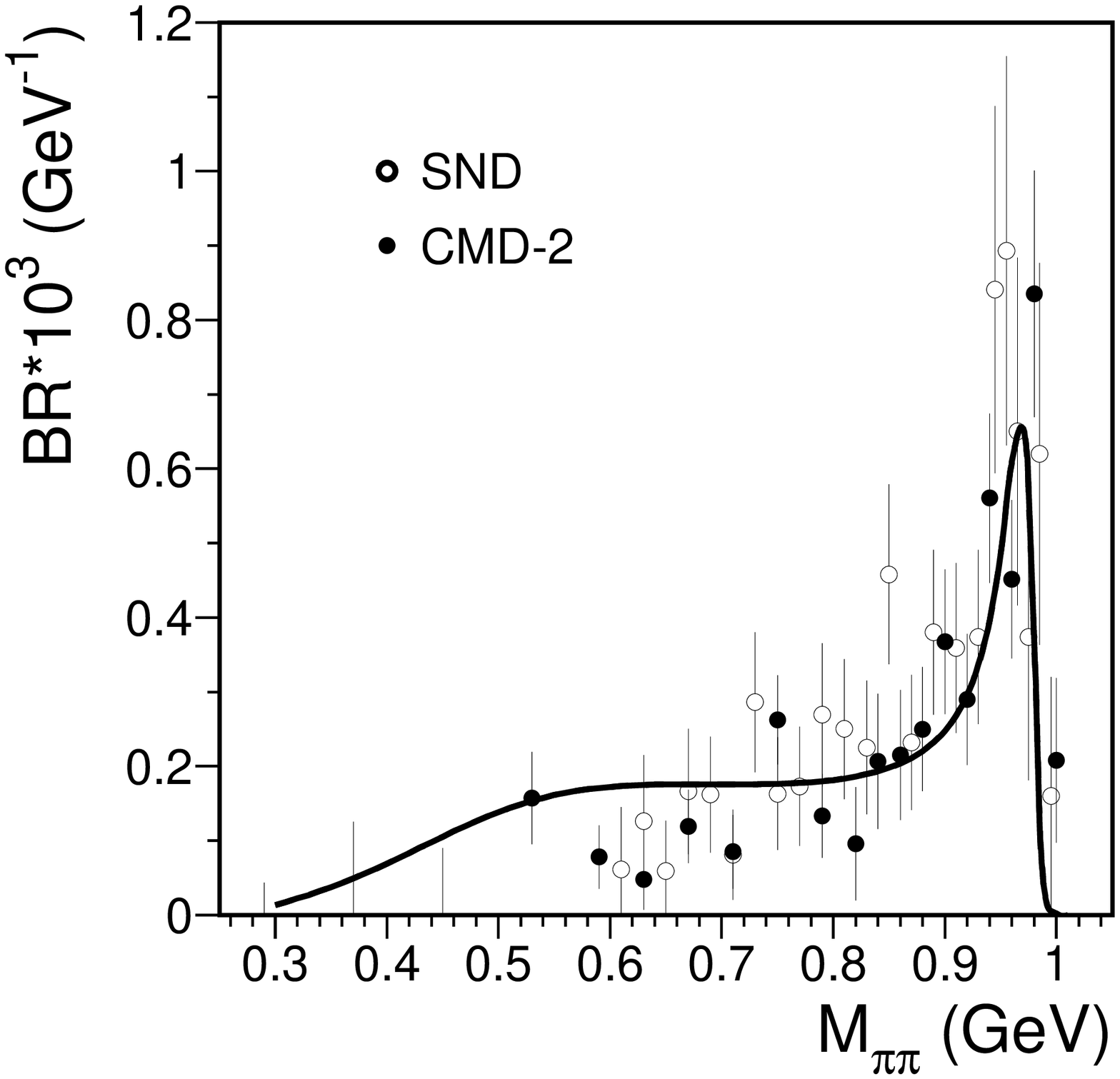}}
} 
\caption{\label{Fig2}
(a) The spectral density of the $q\bar{q}$ state. 
(b) The $\pi\pi$ invariant mass distribution in the
decay $\phi\to\gamma\pi^0\pi^0$.
The curve is the result of our CCM model. 
The experimental points are from \protect\cite{SND98} and \protect\cite{CMD99}. 
} 
\end{figure}

\section{The $\pi\pi - K\bar{K} - q\bar{q}$ Coupled Channel Model}

   Our approach is based on an exactly solvable coupled channel model (CCM), 
which is an extended version of the one studied in \cite{LMZ98};  
it contains the $\pi\pi$, $K\bar{K}$, and $q\bar{q}$ channels 
($J^{PC}=0^{++}$, $I^G=0^+$), 
and the interactions are approximated by separable potentials.   
   The model parameters were determined by fitting the $\pi\pi$ scattering 
amplitude together with the $\pi\pi$ mass distribution in the decay 
$\phi\to\gamma(\pi\pi)_{J=0}$ \cite{SND98,CMD99}. 
The detailed description of the model is given in \cite{Ma00}.   

  The poles of the $S$-matrix in the complex mass plane are shown 
in Fig.\ref{Fig1}. 
  The $f_0(980)$ resonance corresponds to {\it one} $S$-matrix pole 
$M_A=(0.975-i0.017)\;$GeV close to the $K\bar{K}$ threshold; 
this pole has a dynamical origin and represents the molecular-like 
$K\bar{K}$ state.  
  There are two poles corresponding to the $\sigma$ meson: 
$M_B=(0.46-i0.24)\;$GeV on sheet II and $M_D=(0.67-i0.23)\;$GeV on sheet III.  
The dynamical origin of the $\sigma$ meson is related to   
the attractive character of the effective $\pi\pi$ interaction 
which has a partial contribution from the coupling 
via the intermediate scalar $q\bar{q}$ states.   
  Two poles at $M_C=(1.42-i0.26)\;$GeV (sheet II) and 
$M_E=(1.40-i0.25)\;$GeV (sheet III) originate from the bare $q\bar{q}$ state. 
The pole trajectories plotted in Fig.\ref{Fig1} were calculated for the 
$K\bar{K}-\pi\pi$ and $q\bar{q}-\pi\pi$ couplings changing from zero 
(the decoupled channels) to their physical values.      
  
  The mixing between the $q\bar{q}$ and the two--meson channels is 
illustrated in Fig.\ref{Fig2}a showing the calculated spectral density 
$\rho(M)$ of the $q\bar{q}$ state.  
In the weak--coupling limit, there would be a single peak corresponding 
to the bare $q\bar{q}$ state with mass $M_{q\bar{q}}^{(0)}=1.11\;$GeV.       
As a result of the strong $\pi\pi-q\bar{q}$ coupling a substantial fraction 
of the $q\bar{q}$ spectral density is present in the $\sigma$ meson region.       
The peak around $1.3\;$GeV corresponds to the broad resonance originating 
from the $q\bar{q}$ state. 
This peak and the corresponding poles $M_C$ and $M_E$ 
can be associated with the $f_0(1370)$ meson.   

  The molecular picture of the $f_0(980)$ meson was found to be fully 
consistent with the data on the $\phi\to\gamma\pi\pi$ decay.     
Figure~\ref{Fig2}b shows the result of the CCM calculations 
of $\pi\pi$ invariant mass distribution in the decay $\phi\to\gamma\pi^0\pi^0$ 
(see \cite{Ma00} for detailed discussion). 
  The calculated branching ratio 
$BR(\phi\to\gamma\pi^0\pi^0)= 0.5\cdot BR(\phi\to\gamma(\pi^+\pi^-)_{J=0}) = 
1.2\cdot 10^{-4}$ is in good agreement with the experimental data   
$BR(\phi\to\gamma\pi^0\pi^0)=(1.14\pm 0.10 \pm 0.12)\cdot 10^{-4}$ \cite{SND98} 
and  
$BR(\phi\to\gamma\pi^0\pi^0)= (1.08\pm 0.17\pm 0.09)\cdot 10^{-4}$ \cite{CMD99}  
and with the recent calculation in chiral perturbation theory  
$BR(\phi\to\gamma\pi^0\pi^0)=0.8 \cdot 10^{-4}$ \cite{MHOT99}.

\section{Conclusion}

  The structure of the $S$--matrix poles in our coupled channel model 
shows the dynamical origin of the $\sigma$ and $f_0(980)$ mesons: 
both of them are produced by strongly attractive interactions in the $\pi\pi$ 
and $K\bar{K}$ channels.   
  The distinction between genuine $q\bar{q}$ states and dynamical resonances, 
$\sigma$ and $f_0(980)$, can be demonstrated in the limit $N_c\to\infty$, 
which implies vanishing meson--meson interaction, 
where the $q\bar{q}$ states turn into infinitely narrow resonances while the
dynamical states disappear altogether.
  Our calculation of the decay $\phi\to\gamma\pi\pi$ shows that the $K\bar{K}$ 
molecular picture of the $f_0(980)$ is in good agreement with the recent 
experimental data.  

  The structure of the $q\bar{q}$ state embedded into the mesonic continuum
has been analyzed using the calculated $q\bar{q}$ spectral density $\rho(M)$.  
While the high--energy part of $\rho(M)$ is related to the $f_0(1370)$ resonance,   
there is also a significant contribution to $\rho(M)$ in the  
region of $\sigma$ meson, which demonstrates the strong coupling between
the $\pi\pi$ and $q\bar{q}$ channels.   
  The same approach can also be used for studying the QCD sum rules
related to the gluon condensate by including the mixing with the scalar 
glueballs in an extended coupled channel model.  The results of these studies 
will be published elsewhere.


\end{document}